\documentclass[5p,twocolumn,times,number]{elsarticle}
\usepackage{graphicx}
\usepackage{amsmath}
\usepackage{epsfig}
\usepackage{subcaption}
\usepackage{lipsum}
\usepackage{graphicx}

\begin{document}

\begin{frontmatter}

\title{Determination of Position Resolution for LYSO Scintillation Crystals using Geant4 Monte Carlo Code}
\author {M. F. O. Yahya}
\author {F. Kocak\corref{cor1}}
\ead{fkocak@uludag.edu.tr}
\cortext[cor1]{Corresponding author. Tel: +90 2242941709; fax: +90 2242941899.}
\address{Department of Physics, Bursa Uludag University, 16059, Bursa, Turkey}

\begin{abstract}
LYSO scintillation crystals, due to their significant characteristics such as high light yield, fast decay time, small Moli\'ere radius, and good radiation hardness, are proposed to be used for the electromagnetic calorimeter section of the Turkish Accelerator Center Particle Factory (TAC-PF) detector. In this work, the center of gravity technique was used to determine the impact coordinates of an electron initiating an electromagnetic shower in the LYSO array, the calorimeter module contains nine crystals, each 25 mm$\times$25 mm in cross-section and 200 mm in length. The response of the calorimeter module has been studied with electrons having energies in the range 0.1 GeV-2 GeV. By using the Monte Carlo simulation based on Geant4, the two-dimensional position resolution of the module is obtained as $\sigma_{R}(mm)=((3.95\pm0.08)/\sqrt{E})\oplus(1.91\pm0.11)$  at the center of the crystal.
\end{abstract}

\end{frontmatter}

\section{Introduction}
Cerium doped silicate-based heavy scintillation crystals were improved for medical applications in the first place. Later, massive construction abilities of lutetium oxorthosilicate (LSO) \cite{c} and lutetium-yttrium oxorthosilicate (LYSO) \cite{D,T} were found. LYSO crystals have high stopping power $(> 7\, g/{cm^3})$, fast decay time (40 ns) and high light yield (200 times of PWO), and superior radiation hardness against gamma rays, neutrons, and protons \cite{r4,r5,r6}. Furthermore, the crystal emits light in the wavelength region between 360 nm to 600 nm, reaching a peak at 402 nm. Due to all of the above-mentioned properties, this crystal has also attracted the attention of experimental high energy physics research groups working to improve the performance of electromagnetic calorimeters, such as the proposed SuperB forward parts of the endcap calorimeter \cite{r7}, the KLOE experiment \cite{r8}, the COMET experiment at J-PARC \cite{r9}, and the Muon-to-Electron (Mu2e) experiment \cite{r10}. In addition to the lead tungstate (PWO) and Thallium activated Cesium Iodide (CsI(Tl)) crystals, the LYSO scintillation crystals may also be studied for the electromagnetic calorimeter (ECAL) section of the proposed Turkish Accelerator Center-Particle Factory (TAC-PF) detector \cite{r11}. This paper reports the results of a simulation study carried out with Geant4 to evaluate the position resolution of the LYSO scintillation crystals for incident electrons at an energy range from 0.1 GeV to 2 GeV.
\section{Position Resolution of Electromagnetic Calorimeters}

When a particle is sent to the electromagnetic calorimeter, its energy is deposited in the central crystal as well as the crystals around it. The impact position of the particle can be found from the weighted mean of the position of energy deposits in the crystals. 
This technique used to find the position of the incident particle is called the center of gravity method and calculated by:
\begin{equation}
x_{gravity}= \frac{\sum_i x_i E_i}{\sum_i E_i}
\end{equation}

 for the x-coordinate and similarly, 

\begin{equation}
y_{gravity}= \frac{\sum_i y_i E_i}{\sum_i E_i}
\end{equation}
for the y-coordinate. Where $E_i$ is the energy deposited in the $i^{th}$ crystal and $x_i$ and $y_i$ are the coordinates of the $i^{th}$ crystal \cite{r12}. For the ideal position resolution, it is preferable to take the sum over nine crystals due to the effect of large energy fluctuations in the tails of an electromagnetic cascade \cite{r13}. 

\section{Geant4 Simulation and Results} 
Geant4 simulation code \cite{r14} was employed to perform the
simulation process for the electrons passing the electromagnetic calorimeter module consisting of $3\times3$ LYSO scintillation crystals. The electrons were directed perpendicular to the module in the energy range of 0.1 GeV to 2 GeV. Simulation was performed with $\textit{Geant4.10.04-patch-03}$ with $\textit{QGSP-BERT 4.5}$ physics list. The LYSO crystals have a length of 200 mm $(17.5X_0)$ with cross section of 25$\times$25 $mm^2$ $(1.2 R_M)$. In the simulation, in order to obtain the distributions of the center of gravity of the deposited energies in the crystals, electrons were injected at fourteen positions to the central crystal of the LYSO matrix at different energies. The relation between $x_{gravity}$  and  $y_{gravity}$ in the center of the LYSO matrix for 1 GeV electrons can be seen from Figure 1.

\begin{figure}[hbt!]
\centering
\includegraphics[width=0.8\linewidth]{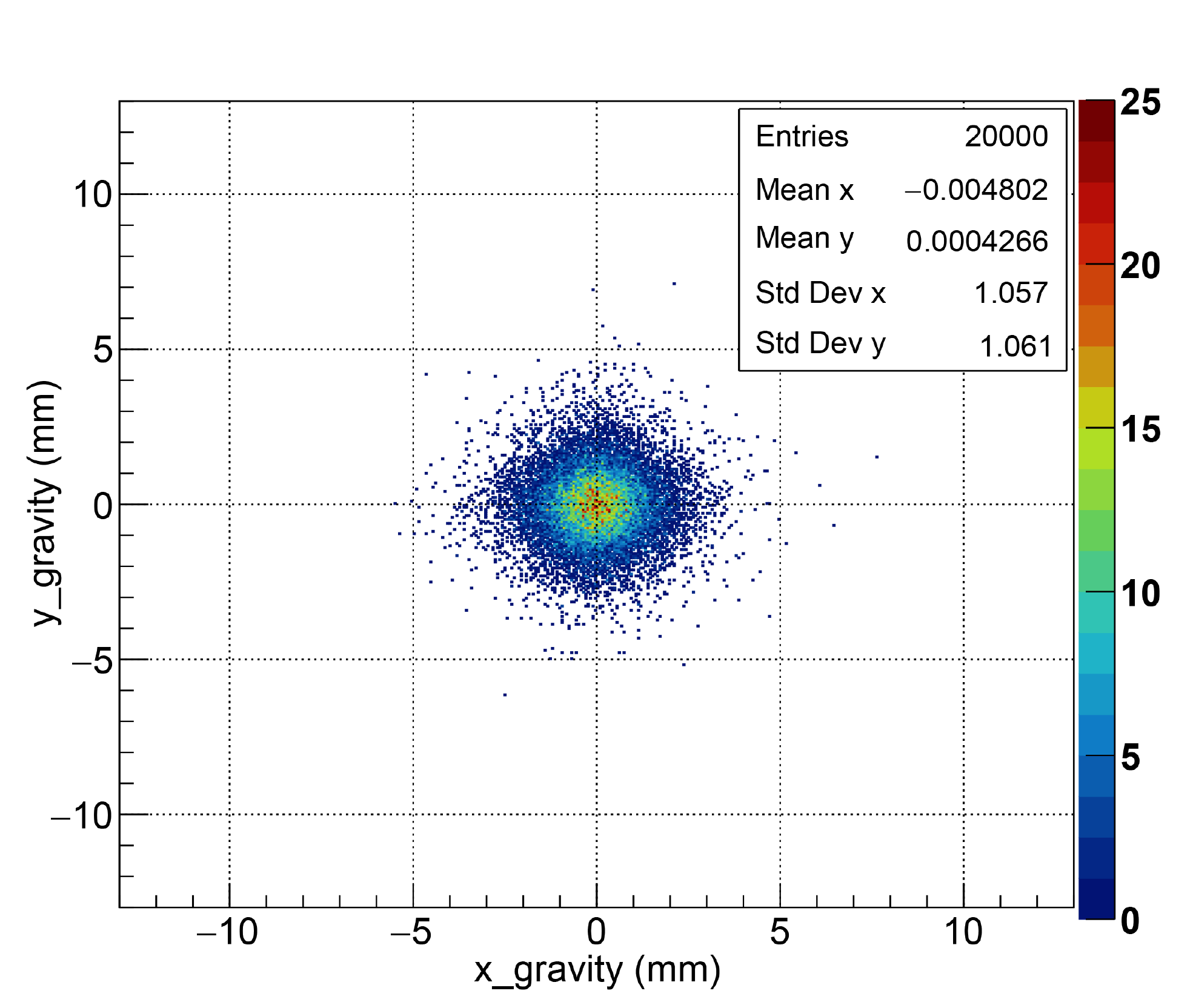} \caption{Relation between $x_{gravity}$ and $y_{gravity}$ in the center of LYSO matrix for 1 GeV electrons.}
\end{figure}

The calculated positions ($x_{gravity}$) versus the true positions ($x_{true}$) are shown in Figure 2 for 1 GeV electrons (S-curve). If the impact point is at the center of the crystal or near the boundary between different crystals, the particle's position can be reconstructed correctly, as can be seen from Figure 2. In other cases, since the most of the energy in the shower is deposited in the hit crystal and there is an exponential fall in the energy shared among neighboring crystals, the position of the particle is systematically miscalculated from equation (1). 

\begin{figure}[hbt!]
\centering
\includegraphics[width=0.8\linewidth]{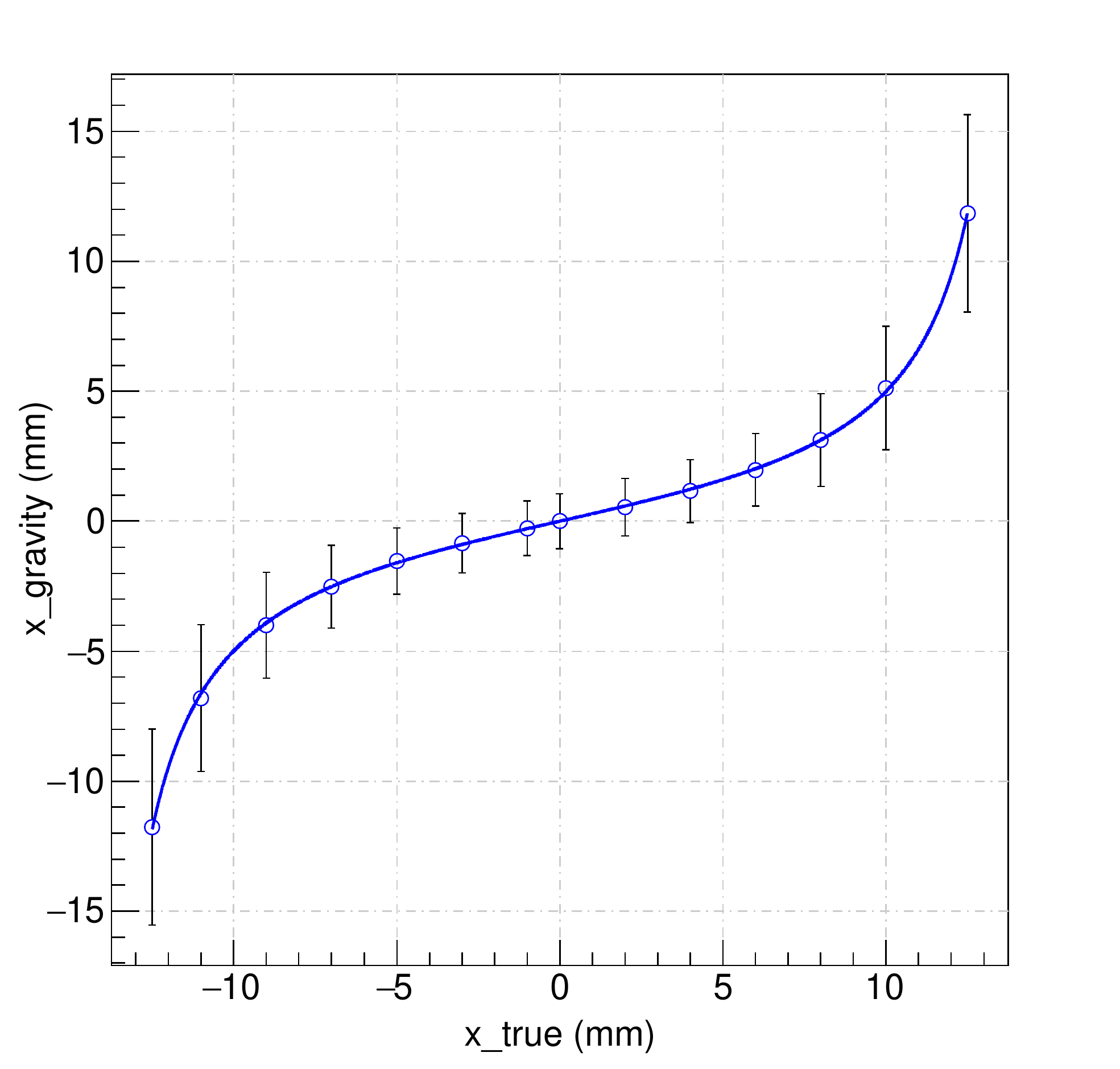} \caption{The $x_{gravity}$ position versus the $x_{true}$ position for 1 GeV electrons. The solid blue line is the S-like curve. }
\end{figure}

\begin{figure}[hbt!]
\centering
\includegraphics[width=0.9\linewidth]{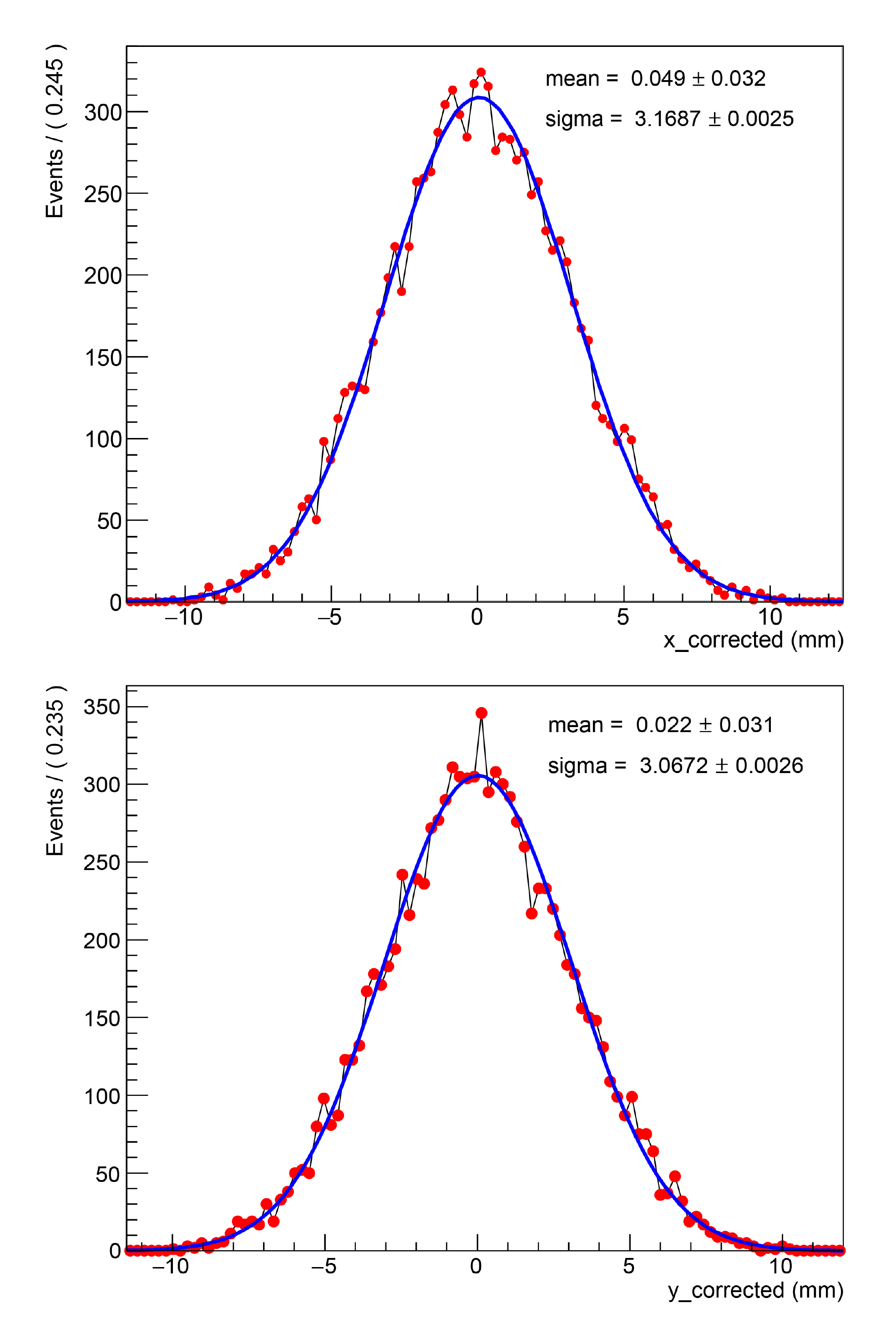} \caption{$x_{corrected}$ and $y_{corrected}$ positions after the S-curve correction, for electron energy of 1 GeV.}
\end{figure}

To remove this non-linear dependence among the true positions $x_{true}$ and the calculated positions $x_{gravity}$, the S-curve fit function was utilized. This function is an empirical algorithm and given by

\begin{equation}
    x_{gravity} =c \tan d (x_{true}-e).
\end{equation}
Here $x_{gravity}$ and $x_{true}$ are presented in mm. As a result of this fit, the values of c, d, and e at 1 GeV are found to be 2.669, 0.108, and 0.001, respectively. Table 1 shows the obtained fit results for the incident electron energies between 0.1 and 2 GeV in the X direction. As the Moli\'ere radius is hardly dependent on energy, the parameters differ slightly depending on the energy of the incident electron. By employing the values c, d, and e which are obtained from the fit, the corrected X position ($x_{corrected}$) is calculated by

\begin{equation}
{x_{corrected}}=\frac{1}{d}\tan^{-1}\frac{x_{gravity}}{c}+e.
\end{equation}

Similar calculations have been also made for the Y direction to obtain the corrected Y position ($y_{corrected}$). As the corrected position distribution spectra have roughly a gaussian shape (see Figure 3), they have been fitted using the gaussian function to obtain the position resolution. The sigmas of these distributions shown in Figure 3 give the calorimeter position resolutions in the X and Y directions for 1 GeV electron energy. Figure 4 shows the corrected position ($x_{corrected}$) versus the true position ($x_{true}$) for 1 GeV electrons in the X direction. Figure 5 displays the position resolutions at the center of the central LYSO crystal (at coordinate x= y= 0) for the incident electron energies from 0.1 GeV to 2 GeV. The Geant4 simulations indicate that the position resolution improves as the energy of the incident electron increases as shown in Figure 5. At the center of the central crystal, the position resolutions depending on the energy of the incident electrons can be parameterized as

\begin{table}[ht]
\caption{Fit parameters for the incident electrons in the X direction.}
\centering
\begin{tabular}{c c c c}
\hline\hline
Electron energy (GeV) & c (mm) & d (rad/mm) & e (mm) \\ [1ex]
\hline
0.1   &	2.166   & 0.112     & -0.007\\
0.25  &	2.327   & 0.110	    &  0.017\\
0.5	  & 2.739	& 0.107 	& -0.003\\
0.75  &	2.702	& 0.108  	& -0.004\\
1     & 2.669   & 0.108 	&  0.001\\
1.25  &	2.637	& 0.108 	& -0.023\\
1.5	  & 2.651	& 0.108 	& -0.003\\
1.75  & 2.654	& 0.108 	&  0.003\\
2	  & 2.613	& 0.108 	& -0.01\\
\hline
\end{tabular}
\end{table}

\begin{figure}[hbt!]
\centering
\includegraphics[width=0.9\linewidth]{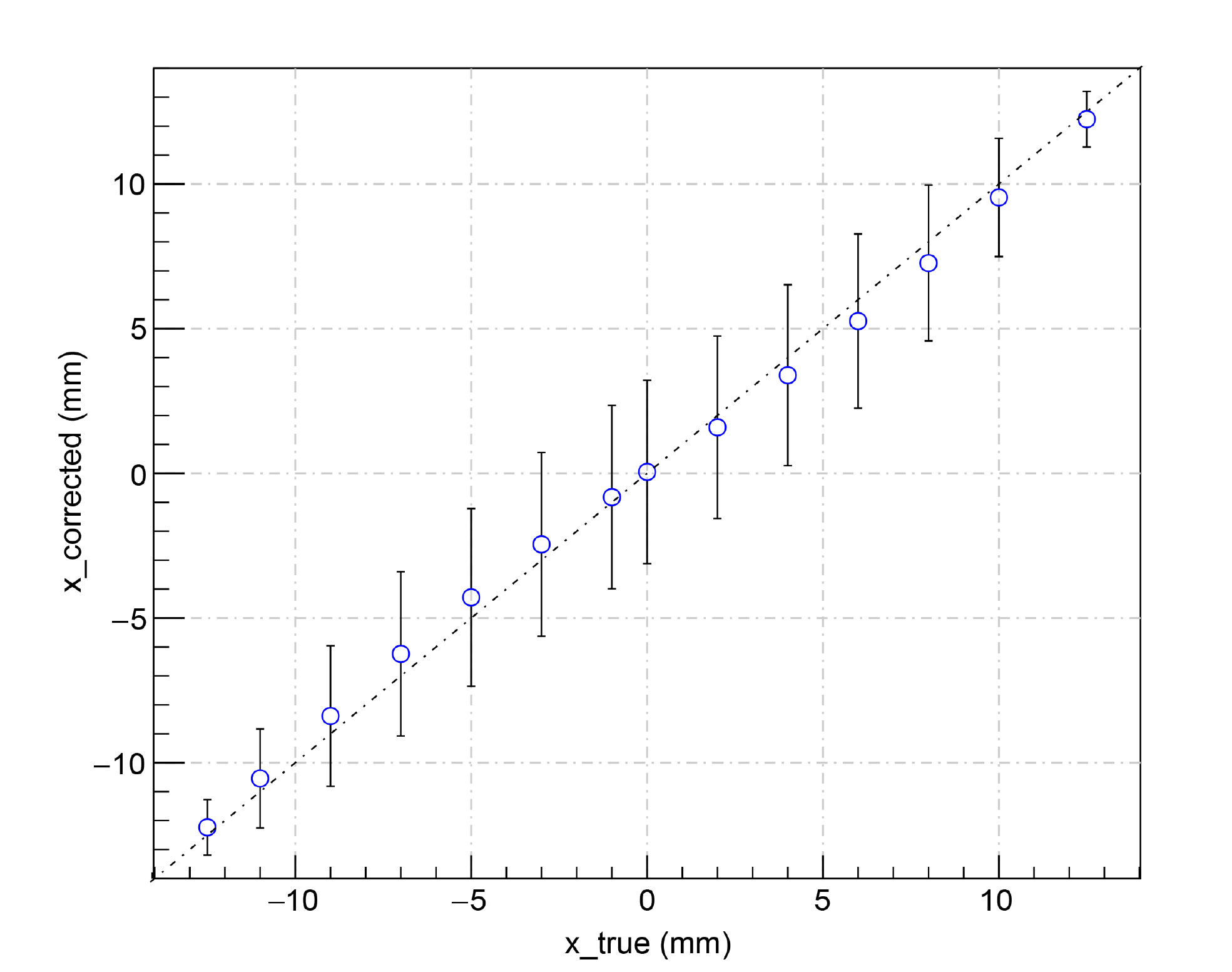} \caption{Dependence of the corrected position $(x_{corrected})$ in the LYSO crystals on the true coordinate $(x_{true})$.}
\end{figure}

\begin{figure}[h]
    \centering
    \includegraphics[width=1.0\linewidth]{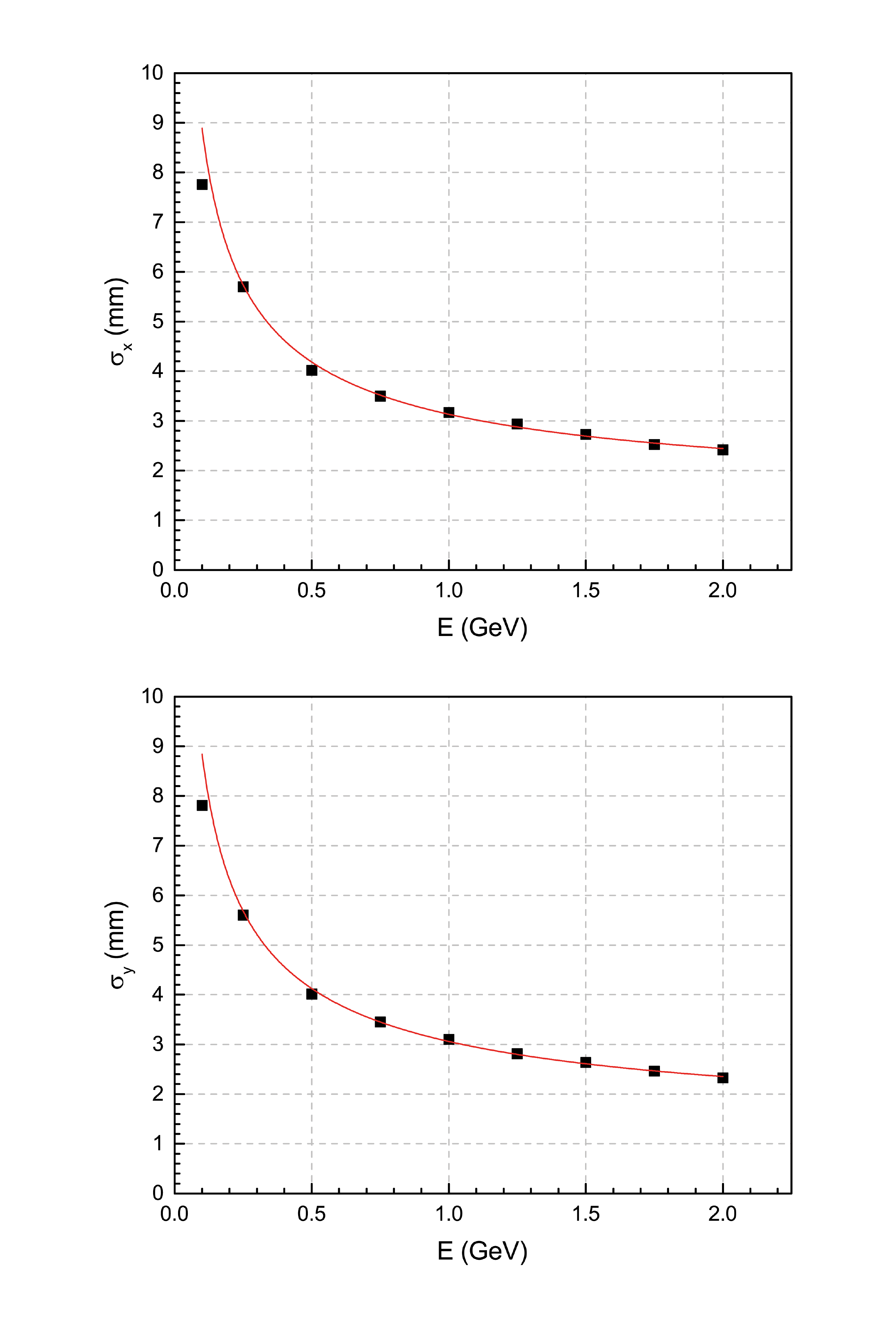}
    \caption{Position resolutions in the X and Y directions at the center of the $3\times3$ LYSO matrix as a function of the energy. The solid lines represent the fits to the datas. The error bars are negligible compared to the symbols shown.}
    \end{figure}

\begin{equation}
    \sigma_x (mm)=\frac{(2.77\pm0.07)}{\sqrt{E}}\oplus(1.46\pm0.10)
\end{equation}
for the X direction and  
\begin{equation}
    \sigma_y (mm)=\frac{(2.77\pm0.05)}{\sqrt{E}}\oplus(1.31\pm0.07)
\end{equation}
for the Y direction.\\

The spatial resolution is not constant on the entire surface of the scintillation crystal and changes depending on the impact position of the incident electron. As can be seen in Figure 6, the simulated position resolution improves towards the edges of the crystal scintillator, and the minimum spatial resolutions are acquired on the edges. This is because the electromagnetic cascade sharing among neighboring crystals starts to become important in that position of the crystal.\\

\begin{figure}[hbt!]
       \centering
       \includegraphics[width=1.0\linewidth]{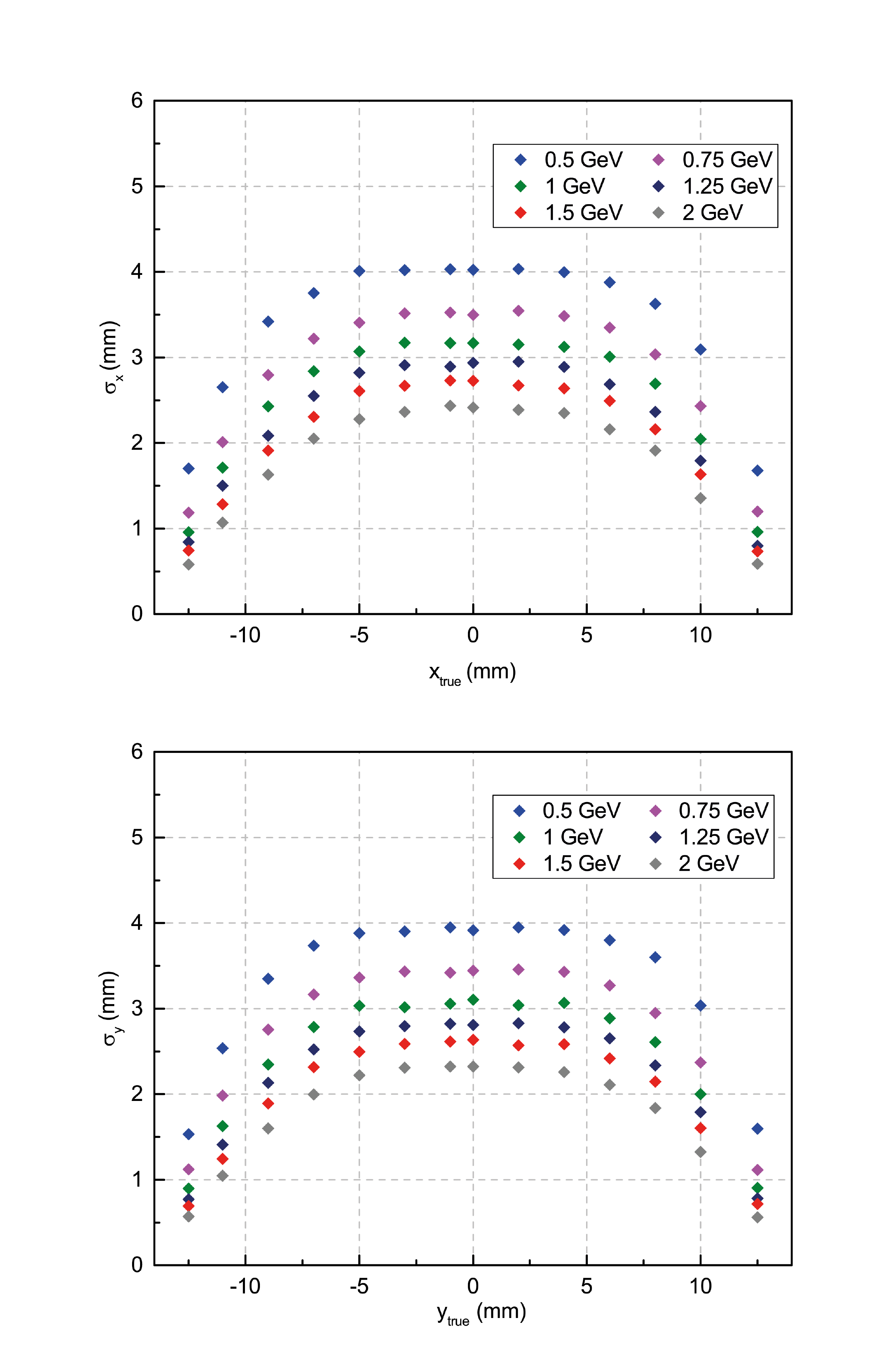}
       \caption{Position resolutions versus the impact position of the electron in the X and Y directions.}
       \end{figure}

One can get two-dimensional position resolution $\sigma_{R}$ by using $\sigma_{x}$ and $\sigma_{y}$ values:
\begin{equation}
    \sigma_R (mm)=\sqrt{\sigma_x^2 + \sigma_y^2}
\end{equation}
Figure 7 shows the two-dimensional position resolution as a function of energy at the center of the $3\times3$ LYSO matrix.
\begin{figure}[hbt!]
\centering
\includegraphics[width=1.0\linewidth]{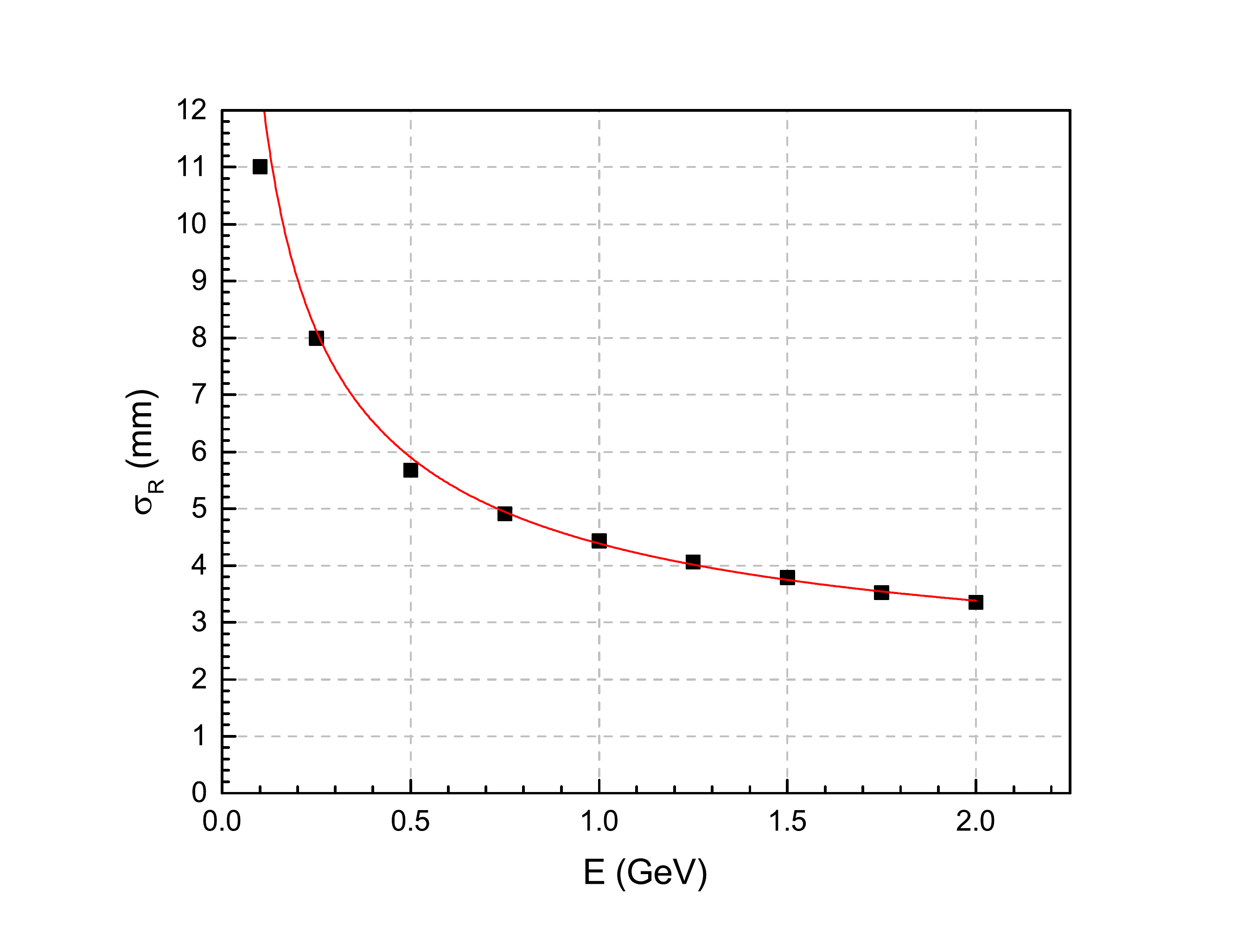}
\caption{Dependence of the two-dimensional position resolution on electron energy.}
\end{figure}

In addition, in order to check the accuracy of the simulation code, a prototype electromagnetic calorimeter developed for the COMET experiment given in Ref \cite{r9} was simulated with the Geant4 code. The mentioned prototype consists of 7x7 LYSO crystals with a dimension of $2\times2\times12$ cm$^{3}$. The position resolution of the prototype was calculated for the incident electrons using the center of gravity technique mentioned above. The simulation results are well-matched with the experimental data with a slight difference at low energies as can be shown in Figure 8. 

\begin{figure}[hbt!]
\centering
\includegraphics[width=1.0\linewidth]{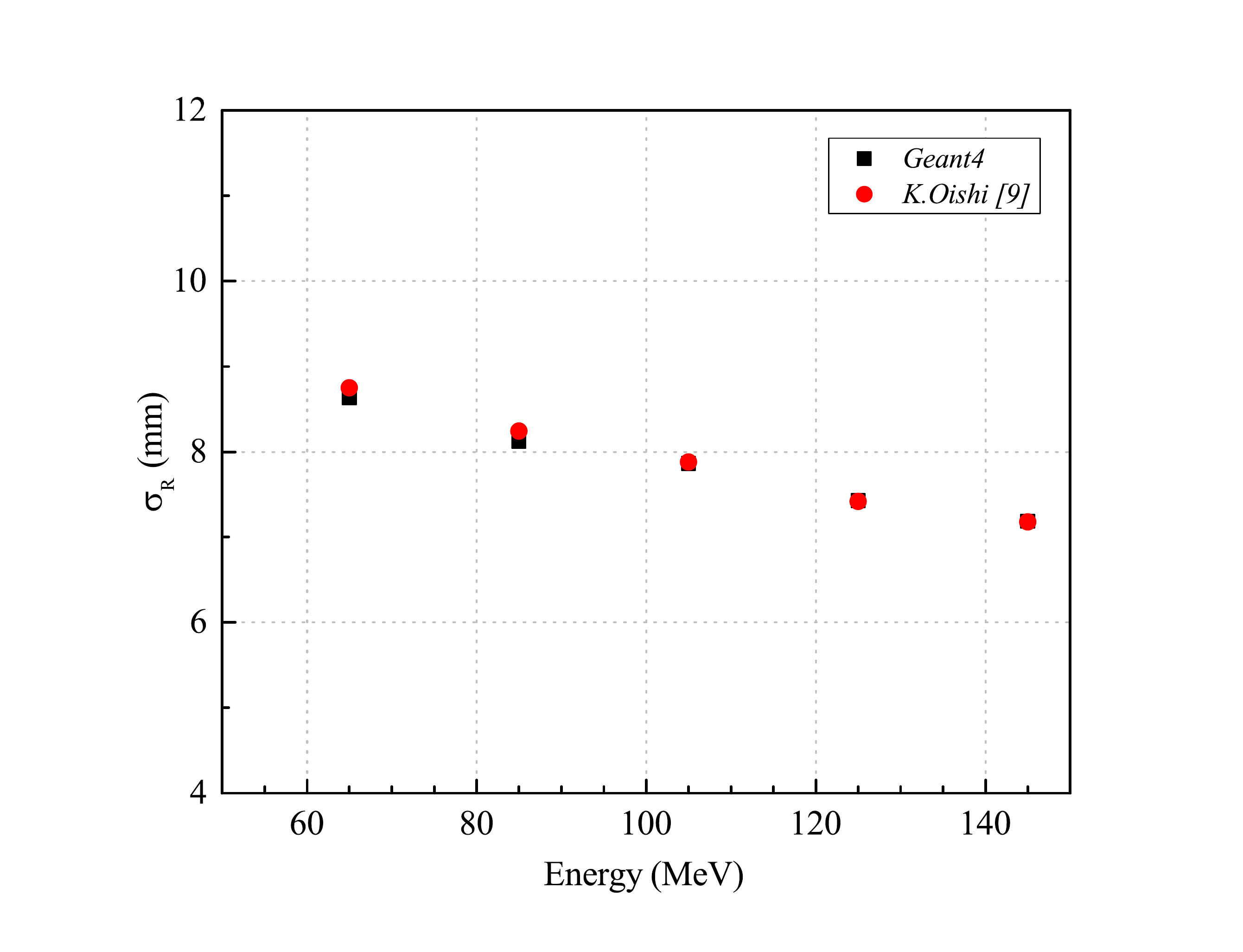} \caption{Comparison of the position resolutions obtained from Ref \cite{r9} with Geant4 simulation for the 7x7 crystal matrix.}
\end{figure}

\section{Conclusion}

The simulation of the position resolution of the electromagnetic calorimeter module made of LYSO crystals for the proposed TAC-PF detector has been performed using Geant4 simulation code. The position resolution was calculated every 2 mm step for the incident electron energies from 0.1 to 2 GeV, as can be shown in Figure 6. The resolution has improved significantly with S-shape correction. 
For electrons entering the calorimeter module in the center of the LYSO crystal, the resolution in the X (Y) direction is 7.76 mm (7.81 mm) at 0.1 GeV and 2.41 mm (2.32 mm) at 2 GeV. The position resolution was calculated as $ \sigma_x (mm)=((2.77\pm0.07)/\sqrt{E})\oplus(1.46\pm0.10)$ for the X direction and a similar result was also obtained for the Y direction. The two-dimensional position resolution was calculated as $\sigma_R(mm)=((3.95\pm0.08)/\sqrt{E})\oplus(1.91\pm0.11)$. Also, by using Function (4) with the fit parameters shown in Table 1 with a minor modification, the impact position of an electron or photon in similar LYSO calorimeters can be deduced from the center of gravity of energy deposits in the crystals.

\section*{Data Availability}
All Figures in the manuscript represent the results obtained by Monte Carlo simulation and the data used to authenticate the findings of this study are cited at relevant places within the text as references.

\section*{Conflicts}
The authors declare that they have no conflicts of interest.

\section*{Acknowledgement}
The numerical calculations reported in this paper were partially performed at TUBITAK ULAKBIM, High Performance and Grid Computing Center (TRUBA resources).

\end{document}